\documentclass[12pt,preprint]{aastex}

\usepackage{psfig,eucal}
\input{epsf}
\shorttitle{..................}

\shortauthors{Pagani Falomo \& Treves}

\begin{document}

\title{Host Galaxies of low z Radio-loud Quasars: A search of HST archives.\footnote{Based
on observations with the NASA/ESA {\em Hubble Space Telescope}, obtained at the Space Telescope Science 
Institute, which is operated by the Association of Universities for Research in Astronomy, Inc.,
under NASA contract NAS 5-26555}}

\author{Claudio Pagani }
\affil{Universit\`a dell'Insubria, via Valleggio 11, 22100 Como, Italy}
\author{ Renato Falomo  }
\affil{Osservatorio Astronomico di Padova, Vicolo dell'Osservatorio 5, 35122 Padova, Italy}
\author{ Aldo Treves }
\affil{Universit\`a dell'Insubria, via Valleggio 11, 22100 Como, Italy}

\begin{abstract}
We searched the HST archives for unpublished WFPC2 images of low
redshift (z $<$ 0.5) radio loud quasars (RLQ). This led to the identification
of 11 objects. We present here the
results of the analysis of these images from which we derive the
properties of their host galaxies.  All objects are clearly resolved and 
their surrounding nebulosity is consistent with an elliptical galaxy model.  These
new data, together with previous published HST observations, form a
sample of 34 sources which significantly expands all previous
studies of low redshift RLQ based on HST data. For this full sample we derive the
average absolute magnitude of the host galaxies 
$<M_R>=-24.01\pm0.48$, and the effective radius
$<R_e>=10.5\pm3.7 $ kpc.
No significant correlation is found between the nucleus and the host 
galaxy luminosity.
Using the relationship between black hole mass ($\CMcal M_{BH}$) 
and bulge luminosity we investigate the relation between $\CMcal M_{BH}$ 
and total radio power for RLQ and compare with other classes
of radio sources. The overall distribution of AGN in the plane $\CMcal M_{BH}$-P$_{radio}$
exhibits a trend for increasing $\CMcal M_{BH}$ with increasing P$_{radio}$ but with a substantial spread.
RLQ occupy the region of most powerful sources and most massive BH.  
The quasars appear to emit over a wide range of power with respect to
their Eddington luminosity as deduced by the estimated $\CMcal M_{BH}$.
\end{abstract}


\section{Introduction}

Extended nebulosities around quasars were
investigated by several groups using various facilities
from the ground (e.g. \citealp{hutch89}, \citealp{veron90}, \citealp{mcleod94}, 
\citealp{taylor96}, \citealp{kotifal00}).  All together these works
have confirmed the presence of the host galaxies.
However their full characterization has been
always problematic because the extreme luminosities of quasars tended
to outshine the surrounding faint nebulosity.  In fact the ability to
measure the global parameters of the host galaxy depends on the shape
of the point spread function (PSF), throughput of the instrument,
also on the distance and on the nuclear-to-host ratio of the sources.
A significant gain to cope with these problems was obtained from
high quality (sub-arcsec) imaging secured in the near-IR where the
contrast between nucleus and host become more favorable.
Nevertheless, the non-homogeneity of ground based data and the 
modest number of studied sources have prevented to produce 
a satisfactory picture of the properties of the quasar hosts.
In particular the issue of distinguishing between elliptical 
and spiral/disc galaxies remained unsettled.

A major impetus in this field came from the observations obtained with
Hubble Space Telescope (HST).  This is because its narrow PSF allows
one to distinguish with unprecedented clearness the host galaxy from
the nucleus.  The advantage is especially important for low redshift
(z $<$0.5) objects where the signal from the host galaxy is
sufficiently high, while at high redshift, in spite of the excellent
PSF, the small throughput of the 2.5m aperture telescope becomes the
main limitation for these studies.  This kind of programs was indeed
among the original scientific motivations of the HST project.

Thus far a number of projects aimed at the study of QSO hosts at low
redshift have been performed with HST (\citealp{bahcall97}; 
\citealp{hooper97}; \citealp{boyce98};   
Kirhakos et al 1999; \citealp{dunlop03}).  These
images indicate that Radio Loud Quasars (RLQs) hosts are 
giant ellipticals, while Radio Quiet Quasars (RQQs)
 are hosted both in ellipticals and spirals.  Although the samples of 
investigated RLQ and RQQ are still small and rarely defined in a
homogeneous way there is a clear tendency for RLQ hosts to be more
luminous (by $\approx$0.5 mag) with respect to the RQQ hosts 
(\citealp{veron90}; \citealp{dunlop03}; \citealp{fal03b}).

Because of the association of RLQ with massive spheroids these sources
have received particular attention since from the properties of their
host galaxy one can estimate the mass of the central
black hole ($\CMcal M_{BH}$) using the relationship between $\CMcal M_{BH}$ and
bulge luminosity (e.g. \citealp{kormendy})  derived from
nearby galaxies with measured BH masses and assuming that 
the relations holds 
unchanged up to z $\sim$ 0.5.

In order to improve the size of the sample and to explore a wider range of properties
of  RLQs we searched the HST archive for unpublished images of low redshift quasars,
finding images for 11 objects. In this work we present the results of the image analysis
of these objects and in particular derive the properties of their host galaxies. 
The new data on RLQ host galaxies are then combined
with previous published results in order to form the largest 
(34 sources) and
homogeneous data set of RLQ hosts investigated by HST. We discuss here 
the properties of the full sample.
In order to be consistent with our previous papers we 
choose H$_0$=50km sec$^{-1}$ Mpc$^{-1}$, q$_0$=0.

\section{Object selection and data analysis}

We carried on a search for non-proprietary  images of RLQ obtained
with WFPC2 from the HST Archive.  To this aim the AGN catalogue of
\citealp{veron} was cross correlated with the HST-WFPC2 pointings using
Multi mission Archive at Space Telescope ({\em MAST}, http://archive.stsci.edu/index.html)
to find out all possible observed sources which were unpublished
at the time of the search or where the issue of the host galaxy
properties was not examined.  We considered only objects with
$z<0.5$, M$_B$ $<$ -23.0  and exposures longer than 200 sec.  The
result of this search produced a total of 11 objects satisfying the
above criteria (see Table 1).  

These objects are distributed in the redshift range z= 0.184 to
z=0.425 (average $<z>$ = 0.346 $\pm$ 0.083).  The distribution of the objects 
over the plane $z$ -- M$_B$ 
is shown in Fig.~1 and compared with previous published objects. The
11 objects span  a range in nuclear luminosity from M$_B$ = -23.1 to
M$_B$ = -26.9.

For each object HST archive observations include typically 3-4 individual images
of the quasar calibrated using the HST-WFPC2 pipeline. These were 
 combined with the IRAF task CRREJ to remove cosmic rays and 
improve the signal to noise ratio.  
The analysis of the QSO images followed the procedures described
 in \citealp{scarpa} for the study of host galaxies of BL Lac objects.  In
particular we have extracted the azimuthally averaged radial
profile of each object after masking any region that are 
contaminated by companion objects.
This procedure offers the advantage to improve significantly the S/N,
allowing to extend the photometry to larger radii but on the
other hand two dimensional information is lost.  Since our main aim
here is to derive the global properties (morphology, luminosity and
scale length) of the host galaxies this approach is 
appropriate.

The key ingredient for the optimal decomposition of the QSO profiles
is the choice of the point spread function (PSF).  It is well known
that while the inner part (radius $<$ 2 arcsec) 
of the PSF can be properly modeled by a
specific software (Tiny Tim, hereafter TT; Krist \& Hook, 1999) 
the external fainter
halo produced by the scattered light cannot. 
Therefore as in \citealp{scarpa} we adopted here a composite PSF,
adding to the TT PSF an additional smooth component 
to take into account the effect of the scattered light.
The latter component gives a significant
contribution to the wings of the PSF at radii larger than 2-3 arcsec.
Since the signal from our targets may extend up to 5 arcsec from the
nucleus, neglecting this contribution for the wings of the PSF may produce
spurious host galaxy images (see an example in \citealp{pian}).
The smooth component is modelled by an exponential law 
and  has been adjusted in such a way to yield a good representation of the radial
brightness profiles of stars observed by WFPC2. 
 Since the original observations refer to different filters and cameras we adopted 
 the above procedure to obtain a representative PSF for each camera/filter combination.
In Figure 2 we show an example of the adopted PSF, compared with 
 that derived solely from TT model and with the average radial profiles of stars.

In Figure 3 we show the images of the quasars after subtraction of a scaled PSF. 
The  normalization of the PSF was set to match the flux 
of the observed target in an annulus of external radius 0.4 arcsec 
and excluding the 
pixels within 0.1 arcsec from the center of each object. 
These PSF subtracted images give an idea of the extended 
component and the immediate environment of the quasars.
In most cases the nebulosity appears smooth and with a 
round shape. No clear signature of spiral structures is 
detected in agreement with previous HST studies of RLQ hosts.

To determine the parameters of the quasar hosts we modelled 
the observed radial brightness profile of each quasar image by  
superposing a central point source (modelled by the PSF) to a
galaxy model (an elliptical or an exponential disk) convolved with its
proper PSF.  The fit  implies three free parameters (normalization of the
PSF and galaxy model plus the scale length of the host galaxy).  The best
decomposition was obtained by a $\chi^2$ minimization algorithm
varying the three parameters (see Table 2).   We
applied this procedure for both models (elliptical and disk galaxies) and found
that in all cases the fit with an elliptical is better or very close
to that of a disk model.  In no case a significantly better fit
with a disk model was obtained. Therefore, in agreement with previous studies
of RLQ and with appearance of the sources in the images, we adopt the
parameters of an elliptical galaxy model in the following discussion.

The results of the fit are reported in Table 2.  Since the observations
pertain to different filters we have transformed all apparent
magnitudes into the R (Cousins) standard band.  To do this we used the
prescription reported by \citealp{holtzman} 
where the magnitudes in the HST filters (F555W, F606W, F702W, F814W) 
are converted into the Johnson-Cousin system in the V, R and I band. 
We also applied a correction ($\Delta m = 0.1$) to make the zero points,  
obtained within  $0.5$ arcsec radius aperture, applicable for surface photometry 
(\citealp{holtzman}. 
To obtain uniform and comparable results apparent magnitudes are then converted 
to R(Cousin) filter assuming V-R=0.61 R-I=0.70
(\citealp{fukugita95}). 

The uncertainty on each parameter has been evaluated from its
variation within the $\chi^2$ map of two parameters and assuming a
$\Delta\chi^2$ = 2.7. The typical uncertainty for the host and
nucleus luminosity is 0.3 mag.  On the other hand, consistently with
previous works (e.g. \citealp{urry00}; \citealp{dunlop03}; 
\citealp{fal03b}) we find that the magnitude of the hosts
are much better constrained than their effective radii, for which the
uncertainty can be as large as 50\%.

\section{Results}

All the quasars in the considered sample (see Table 1) are well resolved.  
In Figure 4 we show for each object the radial brightness profile 
together with its best fit with the elliptical model. 

To check the consistency of our adopted procedure with that of
previous studies using a two dimensional fitting procedure, 
we have analyzed some RLQ observed with HST by \citealp{dunlop03}.
Of the 10 RLQ examined by these authors we randomly chose 5 objects 
(1020-103, 1217+023, 2135-147, 2141+175, 2247+140) 
and found the apparent magnitude of the host galaxy always agrees 
within  $\Delta$m =0.2 and the effective radii within 30\%.

Using the parameters of the fit reported in Table 2 we derived the absolute
magnitudes and effective radii of the host galaxies (see Table 3). 
These absolute magnitudes  have been
corrected for galactic extinction (\citealp{schlegel}), and k-corrected
following \citealp{poggianti}.  
Nuclear luminosities have been k-corrected using the relationship
$2.5*(\alpha-1)*\log(1+z)$ ($\alpha$ = 0.3)  and color corrected
using data from \citealp{cv90}.  

\subsection{The average properties of the host galaxies of RLQ}

In order to investigate the properties of the host galaxies
of low redshift RLQ we 
have combined the new 11 objects discussed in this paper with 
previous data already published in the literature.
The whole data set  has been treated homogeneously in terms of 
cosmology,  k~-~correction, filter and galactic extinction as specified above.
 
To the objects presented here we added the 18 RLQ observed by \citealp{bahcall97}, 
\citealp{boyce98} and \citealp{dunlop03} which were treated homogeneously  
by Falomo et al 2003a. Moreover we included the five RLQ at z $<$ 0.5 studied by 
\citealp{hooper97}. For the latter we used the reported R apparent magnitudes 
and transformed them into absolute quantities following our procedure.                 
 
The final combined sample consists of 34 objects which is up to now the largest sample of 
RLQ observed by HST. The main properties of these sources are given in Table 3. 
Although the sample cannot be considered complete because original selections 
were not homogeneous and in some cases not well defined, we belive that there is no 
significant bias as far as the host galaxy properties are concerned. 
Therefore we are confident that these sources can yield a 
trusty picture of the general properties of RLQ hosts at low redshift. 

We find that the average absolute mag of the full sample of RLQ is 
$<M_R>=-24.01\pm0.48$, and the effective radius $<R_e>$ = 10.5$\pm$3.7 kpc.
The distributions of the absolute magnitudes and of the effective
radii of the host galaxies are shown in Fig 5.
This result confirms that on average the host of RLQ are large and massive ellipticals the 
luminosity of which is in the range between 2 to 5 L$^*$ where L$^*$ is the 
characteristic luminosity of a Schechter function for ellipticals (\citealp{schechter})

\subsection{Host versus nuclear luminosity}

The ratio between the nucleus and host galaxy luminosity in the R band
ranges between 0.3 and 16 with an average value of 3 (see also Figure 5).
A controversial issue concerning a possible relation between 
nuclear and host galaxy luminosities has emerged from the analysis of 
 resolved quasars.
This relation could be  expected if one assumes that the nucleus is emitting at a fixed 
ratio of the Eddington luminosity and that the mass of the central black hole 
is linked with the host luminosity. 
Claims of a positive correlations between the two quantities 
have been proposed by \citealp{hooper97} on the basis of HST 
quasar images of both RLQ and RQQ. However, this was not confirmed by 
other authors using different 
(but still small) samples (e.g. \citealp{dunlop03}).
Moreover it is worth to note that part of the correlation found by \citealp{hooper97} 
is due to the systematic difference of absolute magnitudes between RLQ and RQQ hosts.

Using our data set for 34 RLQ hosts, that covers a wide range of 
data in the plane M$_{nuc}$~-~M$_{host}$, we 
have investigated this issue.
In Figure 6 we report the points for the full sample of RLQ. 
The luminosity of the nucleus is distributed over $\sim$ 4 magnitudes while 
that of the host galaxies remains confined within 1 mag around 3L$^*$.
We do not find a significant correlation between M$_{nuc}$ and M$_{host}$ as 
indicated by the Spearman rank correlation test that yields R$_S$=0.24. 

Using the relationship between $\CMcal M_{BH}$ and the bulge luminosity 
by \citealp{bettoni} we have computed the $\CMcal M_{BH}$ of 
each quasar (and therefore its Eddington luminosity) 
and compared it with the bolometric luminosity of the nucleus.
To transform the observed nuclear luminosity in the R band 
into bolometric luminosity we adopted a fixed correction of 
L$_{bol} \approx 10 \times$ $\lambda$L$_R$ where L$_R$ is the monochromatic 
luminosity of the nuclei in the R band  (\citealp{laor}).
The $\CMcal M_{BH}$ of RLQ covers the interval between $10^{8.6} $ to $10^{9.6} $ 
($<\log (\CMcal M_{BH})>= 9.07\pm0.24$).
In Figure 7 we report the estimated bolometric luminosity of the quasars in 
the sample versus $\CMcal M_{BH}$. 
It turns out that these objects are emitting with a wide range 
of Eddington factor ($\xi=\frac{L}{L_{Edd}}$), from $\xi \approx 10^{-2}$
to values very close ($\xi\approx 1$) to L=L$_{Edd}$. 
The dispersion of $\xi$ is larger than that expected from the uncertainties of 
determination of BH masses .
Therefore if the scatter of the bolometric correction is small, 
this suggests  that the lack of correlation between nucleus and 
host luminosity is due to an intrinsic spread of $\xi$.

\subsection{Relation between $\CMcal M_{BH}$  and radio power}

Basing on a small number of nearby galaxies with known BH masses it was
suggested by \citealp{franceschini} that $\CMcal M_{BH}$ scales
with the total radio luminosity $L_{radio}$ at 5~GHz ($L_{radio}$
$\sim$ $\CMcal M_{BH}^{2.5}$). This correlation appears to hold over
$\sim$3 order of magnitudes for $\CMcal M_{BH}$ and, given its
steepness, it was proposed as a tool to predict $\CMcal M_{BH}$ from
the radio flux.  
A similar link between $L_{radio}$ and $\CMcal M_{BH}$ in various types
of active galaxies was claimed  by \citealp{laor01} and \citealp{lacy}.
However, a recent analysis by \citealp{ho} of this relationship  
indicated that $\CMcal M_{BH}$ is only loosely related with the radio power.  

In Figure~\ref{Fig8} we show the data for our sample of RLQs in the plane
$\CMcal M_{BH}-\log P_{5GHz}$(total) together with 
other samples of galaxies at
various levels of nuclear activity investigated by \citealp{ho} and
\citealp{oshlack} and low redshift radio-galaxies (\citealp{bettoni}).  
For RLQs we found no correlation between the two quantities
(Spearman correlation coefficient R$_S$=0.21). 
The overall distribution of the AGN in the plane $\CMcal M_{BH}$-P$_{rad}$
exhibits a trend of larger radio power for increasing  $\CMcal M_{BH}$
but with a very large spread (4--5 orders of magnitude). 
The FSRQ data (\citealp{oshlack}), which appear to deviate from the overall plot, 
may become consistent
with the above trend if the nuclear flux is corrected for 
Doppler boosting. Since these BH masses have been inferred from H$_\beta$ width the 
possible effects of orientation, related to a disk geometry of the 
line emitting region, could shift the FSRQ points in Figure 8 to the right 
(\citealp{jarvis}).

The new data points for RLQ together with those of low z RG strengthen
the suggestion (\citealp{dunlop03}) that radio sources of various classes are 
encompassed by two limits in the $P_{tot}$-$\CMcal M_{BH}$ plane (see Fig~\ref{Fig8}).

\section{Conclusions}

We have presented the results from the analysis of 11 unpublished 
WFPC2 images of low redshift RLQ retrieved from HST archive.
These data, together with previous HST observations allow us 
to construct a homogeneous data set of 34 RLQ for z$<$0.5.

The main conclusions of this study are:

1) all RLQ are hosted in massive elliptical galaxies with 
luminosity in the range 2 to 5~L$^*$. 
The average absolute magnitude in the R band $<M_R>=-24.01\pm0.48$.

2) converting the host galaxy bulge luminosity into BH masses  
it is found that RLQ contain supermassive BH of 10$^{8.6}$ to 10$^{9.6}$ $M_\odot$ 

3) no significant correlation is present between the host and 
nucleus luminosities; 
this suggests that the nuclei are emitting in a large range of regimes, 
from 1\% to $\sim$100\% of their Eddington luminosity. 

4) the radio power of RLQ is not correlated with the central BH mass; 
the objects occupy the region of most powerful sources and most massive BH in the
overall trend of AGN in the $P_{rad}$- $\CMcal M_{BH}$ plane.

\section*{Acknowledgments}
This work was partly supported by the Italian Ministry for University 
and Research (MURST) under COFIN 2001/028773,
ASI-IR 115 and ASI-IR 35, ASI I/R/086/02. 
This research has made use of the NASA/IPAC Extragalactic Database 
{\em(NED)} which is operated by the 
Jet Propulsion Laboratory, California Institute of Technology, 
under contract with the National Aeronautics and Space Administration.

%TABLES IN DELUXETABLE FORMAT

\begin{deluxetable}{l l c c c c c c c c}
\tablecolumns{10}
\tablewidth{0pc} 
\tablecaption {TABLE 1}
\tablecaption  {Journal of the observations}
\tablehead{
 \colhead{Name} & \colhead{Alias} & \colhead{z}  & \colhead{RA}  & \colhead{DEC} &
 \colhead{${V_{QSO}}$\tablenotemark{a}} & \colhead{Filter}  &  \colhead{CAM\tablenotemark{b}} & 
 \colhead{T$_{exp}$} & \colhead{PI\tablenotemark{c}} \\
 & & &(h m s)& (d m s ) &  &       &        & (s) & }
\startdata 
0110+297 &            &  0.363 & 01 13 24 & +29 58 15  &  17.0   & F814W &  WF2 &  1900  &  DU   \\
0133+207 & 3C47       &  0.425 & 01 36 24 & +20 57 27  &  18.1   & F555W &  PC1&   600  &  SP   \\
0340+048 & 3C93       &  0.357 & 03 43 30 & +04 57 49  &  18.14  & F555W &  PC1&   600  &  SP   \\
0812+020 &            &  0.402 & 08 15 22 & +01 54 59  &  17.10  & F814W &  WF2 &  1200  &  DU   \\
0837-120 & 3C206      &  0.198 & 08 39 50 & -12 14 34  &  15.76  & F702W &  PC1&   600  &  SP   \\
0903+169 & 3C215      &  0.412 & 09 06 31 & +16 46 11  &  18.27  & F814W &  PC1&  5000  &  EL   \\ 
1058+110 & 4C10.30    &  0.423 & 11 00 47 & +10 46 13  &  17.10  & F814W &  WF2 &  1200  &  DU   \\
1100+772 & 3C249.1    &  0.315 & 11 04 13 & +76 58 58  &  15.72  & F555W &  PC1&   600  &  SP   \\
1232-249 &            &  0.355 & 12 35 37 & -25 12 17  &  17.18  & F814W &  WF2 &  1800  &  DU   \\
1309+355 &            &  0.184 & 13 12 17 & +35 15 21  &  15.64  & F606W &  WF3 &   200  &  BA   \\
1512+370 &            &  0.370 & 15 14 43 & +36 50 50  &  16.27  & F814W &  PC1&   640  &  ST   \\	
\enddata
\label{tab4}
\tablenotetext{a} {QSO magnitudes in V filter from the \citealp{veron} catalogue;}
\tablenotetext{b} {Position of the QSO on the WFPC2: PC1=Planetary Camera, WF2=Camera 2, 
WF3=Camera 3;}
\tablenotetext{c} {Principal Investigator (PI) of the HST program: DU=Dunlop, J.S.; 
EL=Ellingson, E.; SP=Sparks, W.B.; ST=Stockton, A.; BA=Bahcall, J.N.}
\end{deluxetable}

\begin{deluxetable}{l c c c c c c c}
\tablecolumns{8}
\tablewidth{0pc} 
\tablecaption {TABLE 2}
\tablecaption  {Results of the fit of RLQ images}
\tablehead{
 \colhead{Name} & \colhead{z}  & \colhead{R$_{host}$\tablenotemark{a}}  & \colhead{Re} &
 \colhead{R$_{nucl}$\tablenotemark{b}} & \colhead{kcorr\tablenotemark{c}}  &  \colhead{Ar\tablenotemark{d}} & \colhead{${\chi^2_{Disk} / \chi^2_{Ell}}$\tablenotemark{e}} \\
   &    &  & (arcsec)&   &      &      & }
\startdata 
0110+297  & 0.363 & $ 18.86\pm  0.15 $   & $ 1.80  \pm  0.50 $ &17.01  & 0.24   & 0.12  &  3.72\\
0133+207  & 0.425 & $ 19.66\pm  0.50  $   & $ 0.95 \pm  0.75$ &17.73  & 1.42   & 0.20  &  0.97 \\
0340+048  & 0.357 & $ 19.82\pm  0.40  $   & $ 1.25 \pm  0.45$ &18.44  & 1.11   & 0.79  &  1.30 \\
0812+020  & 0.402 & $ 18.41\pm  0.15 $   & $ 1.60  \pm  0.50 $ &16.21  & 0.28   & 0.06  &  3.04\\
0837-120  & 0.198 & $ 17.68\pm  0.20  $   & $ 2.30  \pm  0.70 $ &16.09  & 0.22   & 0.11  &  1.43 \\
0903+169  & 0.412 & $ 18.94\pm  0.25 $   & $ 0.95 \pm  0.25$ &17.30  & 0.29   & 0.08  &  2.60 \\ 
1058+110  & 0.423 & $ 19.62\pm  0.30  $   & $ 2.10  \pm  1.00 $ &17.53  & 0.30   & 0.05  &  1.66\\
1100+772  & 0.315 & $ 17.73\pm  0.50  $   & $ 1.25 \pm  0.65$ &15.72  & 0.39   & 0.11  &  0.90 \\
1232-249  & 0.355 & $ 19.12\pm  0.30  $   & $ 1.40  \pm  0.75$ &16.16  & 0.24   & 0.19  &  1.40 \\
1309+355  & 0.184 & $ 16.40\pm  0.25 $   & $ 1.10  \pm  0.25$ &15.90  & 0.38   & 0.02  &  7.82 \\
1512+370  & 0.370 & $ 18.73\pm 0.45  $   & $ 1.20  \pm  0.75$ &16.09  & 0.25   & 0.04  &  0.97 \\	
\enddata
\tablenotetext{a}{Host apparent magnitude in R(Cousins) filter, adapted from {\em WFPC2} filters using
\citealp{holtzman} calibration (see text);}   
\tablenotetext{b} {Nuclear apparent magnitude in R band;}
\tablenotetext{c} {K-correction of the host galaxy from \citealp{poggianti};} 
\tablenotetext{d} {Galactic reddening from \citealp{schlegel};}
\tablenotetext{e} {$\chi^2$ ratio from best fit minimization for a model with disk host and with elliptical host.}
\label{tab2}
\end{deluxetable}

%table 3: QSO AND HOST PROPERTIES

\begin{deluxetable}{l c c c c c c}
\tablecolumns{7}
\tablewidth{0pc} 
\tablecaption {TABLE 3}
\tablecaption  {QSOs and host galaxies properties}
\tablehead{
 \colhead{Name} & \colhead{z}  & \colhead{M$_R$(host)\tablenotemark{a}}  & \colhead{Re\tablenotemark{b}} &
 \colhead{M$_R$(nucl)} & \colhead{LogP$_{5GHz}$}  &  \colhead{$L_{nucl} / L_{host}$} \\
   &    &  & (kpc)&   &   &   }
\startdata 
0110+297$ ^*$  & 0.363  &  -23.56   & 13.4   & -24.82   &  26.40     &   3.19  \\	
0133+207$ ^*$  & 0.425  &  -24.41   & 15.5   & -24.91   &  27.20     &   1.59  \\	
3C48           & 0.367  &  -25.08   &   8.1  & -26.08   &  27.67     &   2.51  \\
PHL1093        & 0.258  &  -24.34   &  15.0  & -23.82   &  26.38     &  0.62  \\
PKS0202-76     & 0.389  &  -23.30   &   4.9  & -24.20   &  26.97     &   4.83  \\
0312-77        & 0.223  &  -24.28   &  17.7  & -23.84   &  26.25     &   0.67  \\
0340+048$ ^*$  & 0.357  &  -24.09   &  5.9   & -24.13   &  26.85     &   1.04  \\	
0736+017       & 0.191  &  -24.18   &  13.3  & -24.50   &  26.53     &   1.34  \\ 
0812+0202$ ^*$ & 0.402  &  -24.24   & 11.4   & -25.95   &  26.85     &   4.83  \\	
0837-120$ ^*$  & 0.198  &  -23.23   & 13.1   & -24.46   &  26.14     &   3.10  \\	
0903+169$ ^*$  & 0.412  &  -23.80   &  6.9   & -24.94   &  26.64     &   2.86  \\	
1004+130       & 0.240  &  -24.34   &   8.2  & -25.98   &  26.13     &   4.53  \\
1020-103       & 0.197  &  -23.68   &   7.1  & -23.75   &  25.94     &   1.07  \\ 
1058+110$ ^*$  & 0.423  &  -23.17   & 15.4   & -24.71   &  26.47     &    4.13  \\             
1100+772$ ^*$  & 0.315  &  -24.47   & 10.0   & -26.22   &  26.65     &   5.01  \\
1138+000       & 0.500  &  -24.36   &  ***   & -24.69   &  ****      &   1.35  \\
1217+023       & 0.240  &  -24.08   &  11.1  & -24.64   &  26.34     &   1.67  \\ 
1218+175       & 0.444  &  -23.66   &  ***   & -23.59   &  25.81     &  0.94  \\     	      
1222+125       & 0.415  &  -24.20   &  ***   & -24.36   &  25.74     &   1.15  \\  
3C273          & 0.158  &  -24.39   &   8.5  & -27.40   &  27.74     &   16.00  \\ 
1230-0015      & 0.470  &  -24.50   &  ***   & -24.97   &  26.11     &   1.54  \\
1232-249$ ^*$  & 0.355  &  -23.31   &  8.2   & -25.70   &  26.75     &   9.04  \\	           
1302-102       & 0.286  &  -23.91   &   7.0  & -26.47   &  26.55     &   10.57  \\ 
1309+355$ ^*$  & 0.184  &  -24.41   &  5.4   & -24.70   &  24.85     &   1.301  \\	
B21425+267     & 0.366  &  -24.37   &  10.1  & -26.49   &  25.96     &   7.06  \\ 
1512+370$ ^*$  & 0.370  &  -23.66   &  8.2   & -25.61   &  26.48     &   6.03  \\   
3C323.1        & 0.264  &  -23.33   &   8.2  & -24.82   &  26.51     &   3.95  \\       
3C351          & 0.371  &  -24.67   &   7.8  & -25.33   &  27.02     &   3.84  \\       
2135-147       & 0.200  &  -23.50   &  11.6  & -24.40   &  26.45     &   2.29  \\      
OX169          & 0.213  &  -24.04   &   8.2  & -25.00   &  26.41     &   2.42  \\      
2247+140       & 0.237  &  -24.19   &  13.5  & -24.09   &  26.56     &   0.91  \\      
2349-014       & 0.173  &  -24.51   &  19.2  & -24.21   &  ****      &  0.75  \\     
2351-0036      & 0.460  &  -23.33   &  ***   & -24.53   &  26.73     &    3.02  \\
2355-082       & 0.210  &  -23.97   &  10.4  & -23.29   &  25.58     &  0.53  \\
\enddata
\tablenotetext{*} {Objects analyzed in this work; }
\tablenotetext{a} {Absolute magnitues are in R(Cousin) filter, k-corrected 
and corrected for galactic extinction using \citealp{schlegel};}
\tablenotetext{b} {\citealp{hooper97} do not report values of the effective radius. }
\label{tab3}
\end{deluxetable}

% FIG 1  - z distribution
\begin{figure}[h]
\plotone{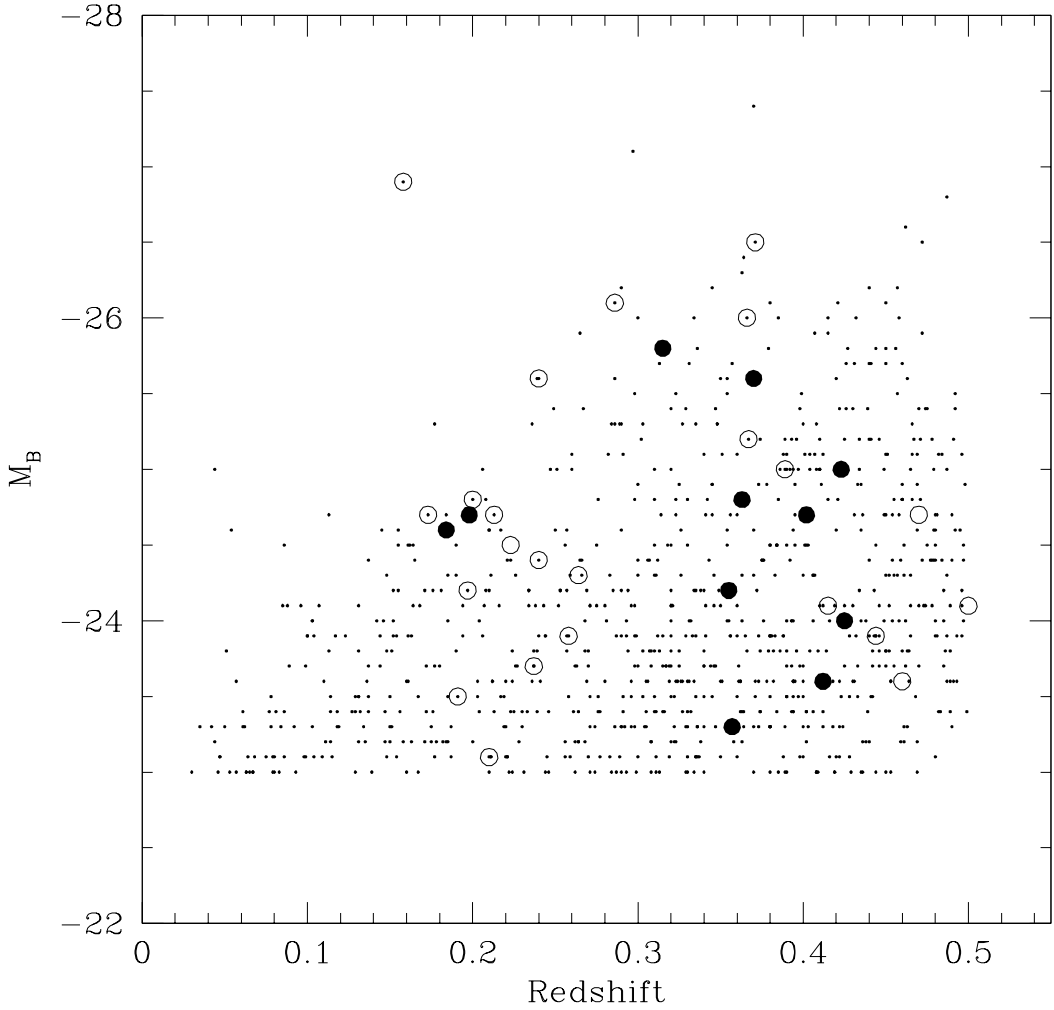}
\caption{Distribution of the RLQ in the sample over the plane M$_B$ -- redshift. 
New 11 objects (filled circles) and previously published sources  (open circles) 
investigated with HST and 
WFPC2 compared with the QSO in the \citealp{veron} catalogue (small points).}
\end{figure}

% FIG 2 - PSF
\begin{figure}[h]
\plotone{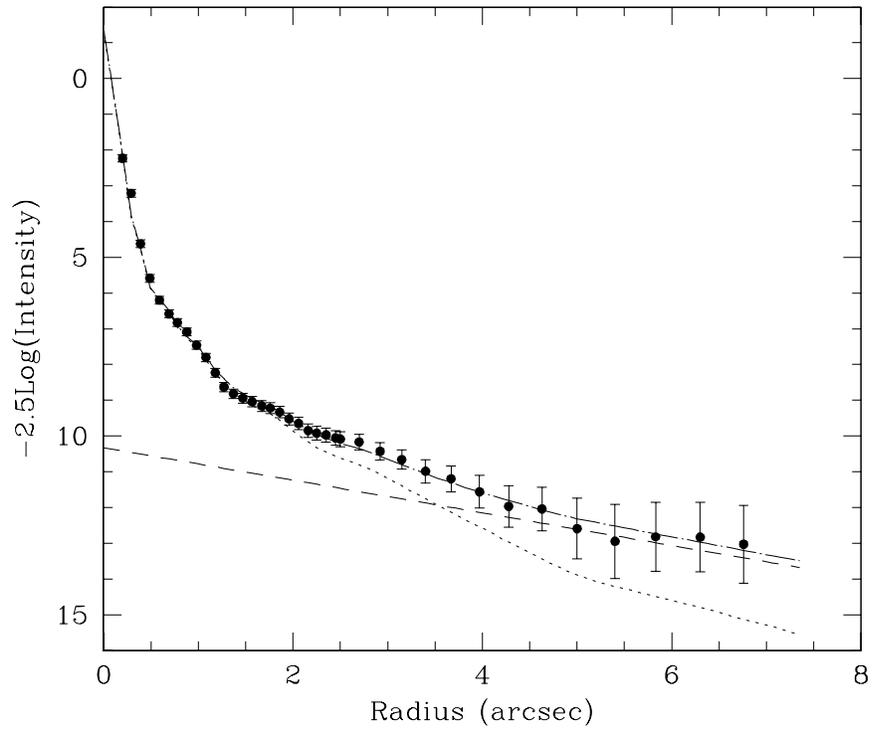}
\caption{The adopted PSF (solid-dot line) is a combination of the TinyTim PSF (dotted line) with an exponential 
component (dashed line) to take into account the scattered light (see text).
Data points represent the observed averaged radial profiles of stars.}
\end{figure}

% FIG 3 - images
\begin{figure}[h]
\epsscale{0.90}
\plotone{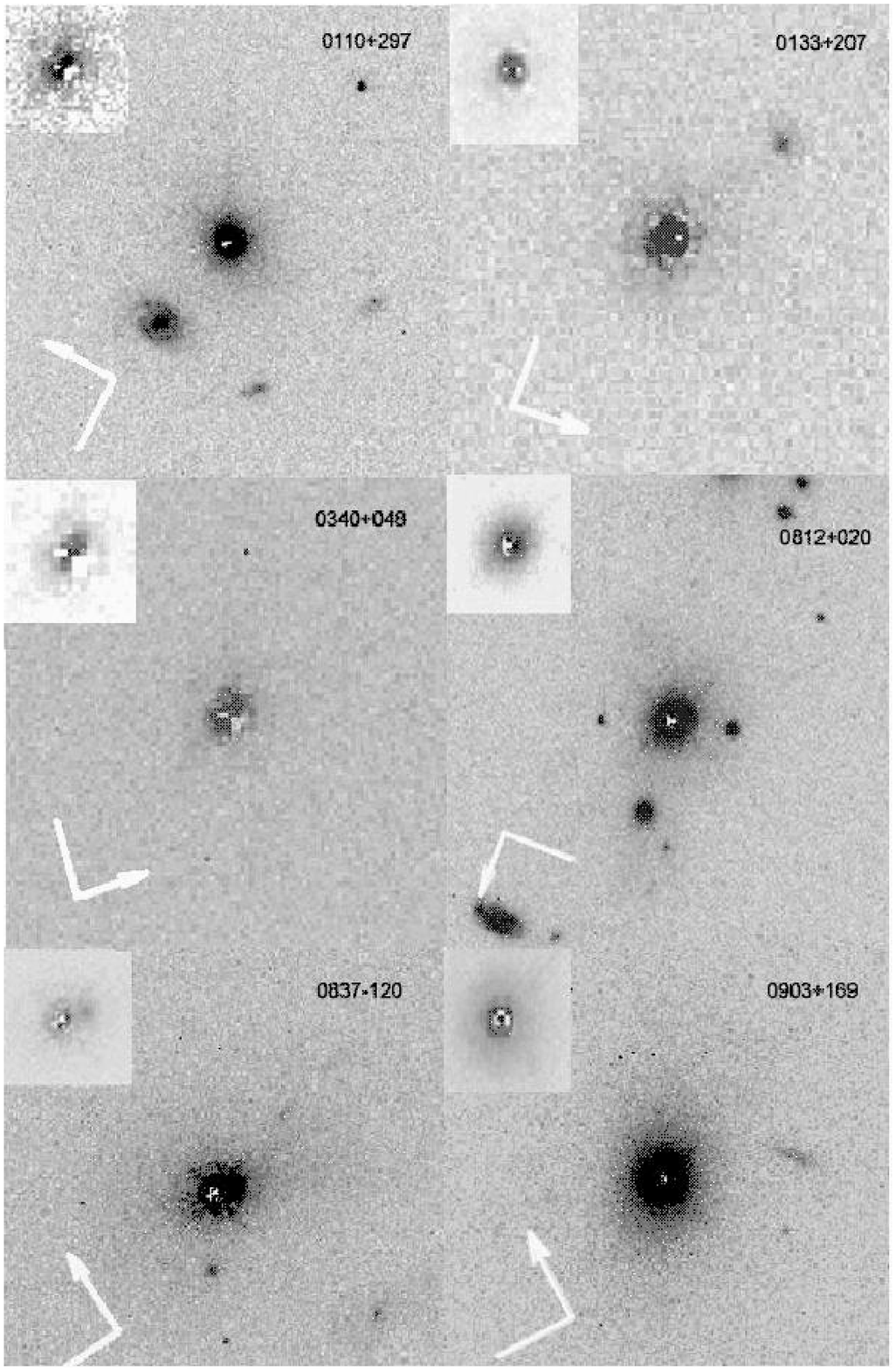}
\caption{Portion of the WFPC2 QSO images after subtraction of the PSF component.
In the inset  of each panel the inner part of the quasar  is shown in a logarithmic
grey-scale. For each field the North direction is represented by the arrow. 
East is on the left of the arrow.  The field of view shown in each panel is approximately 
15$\times$ 15 arcsec.}
\end{figure}
\addtocounter{figure}{-1}%

% FIG 3 images
\begin{figure}[h]
\epsscale{0.9}
\plotone{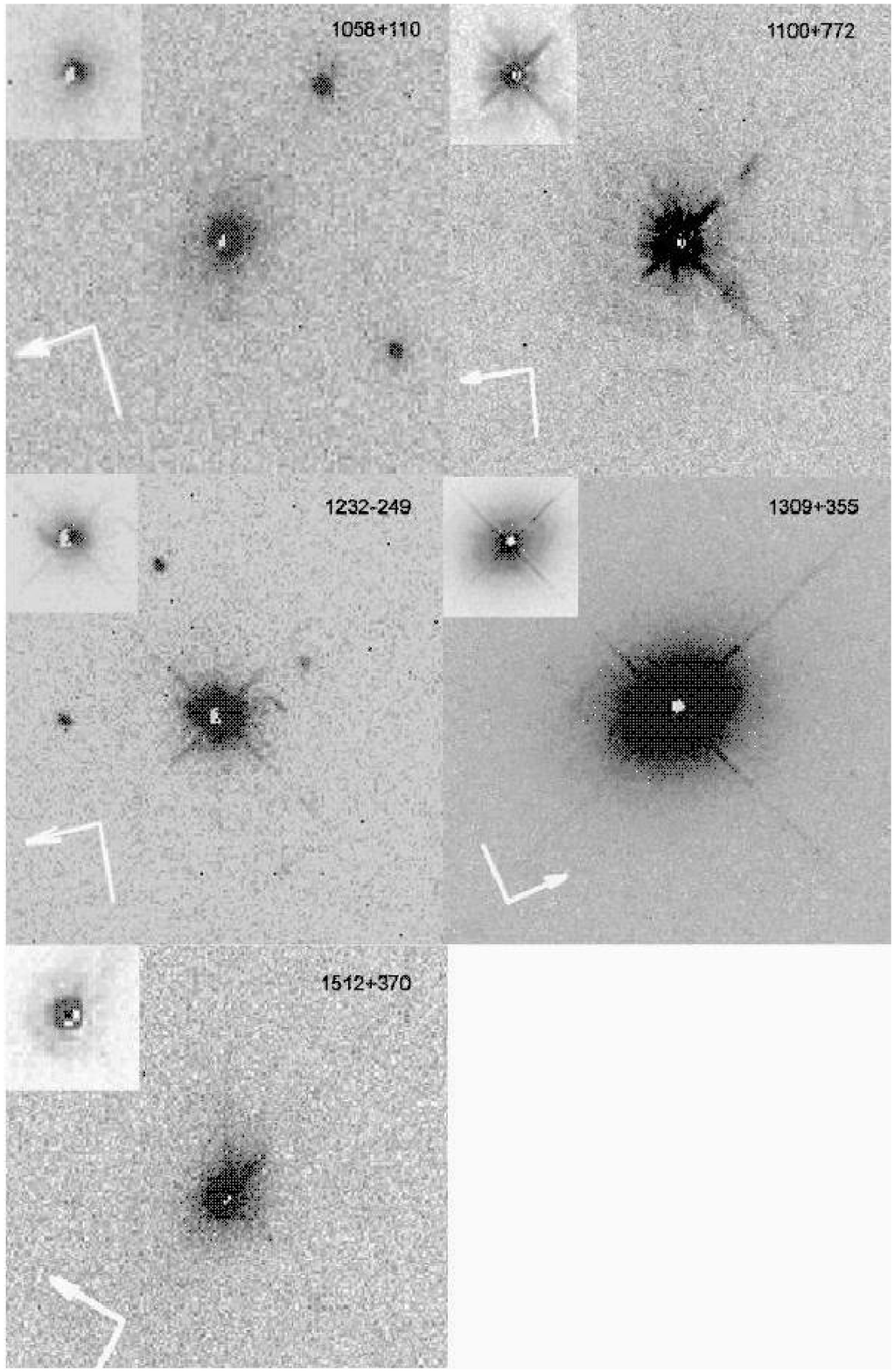}
\caption{continued.}
\end{figure}

% FIG 4 radial profiles
%
\begin{figure}[h]
\plotone{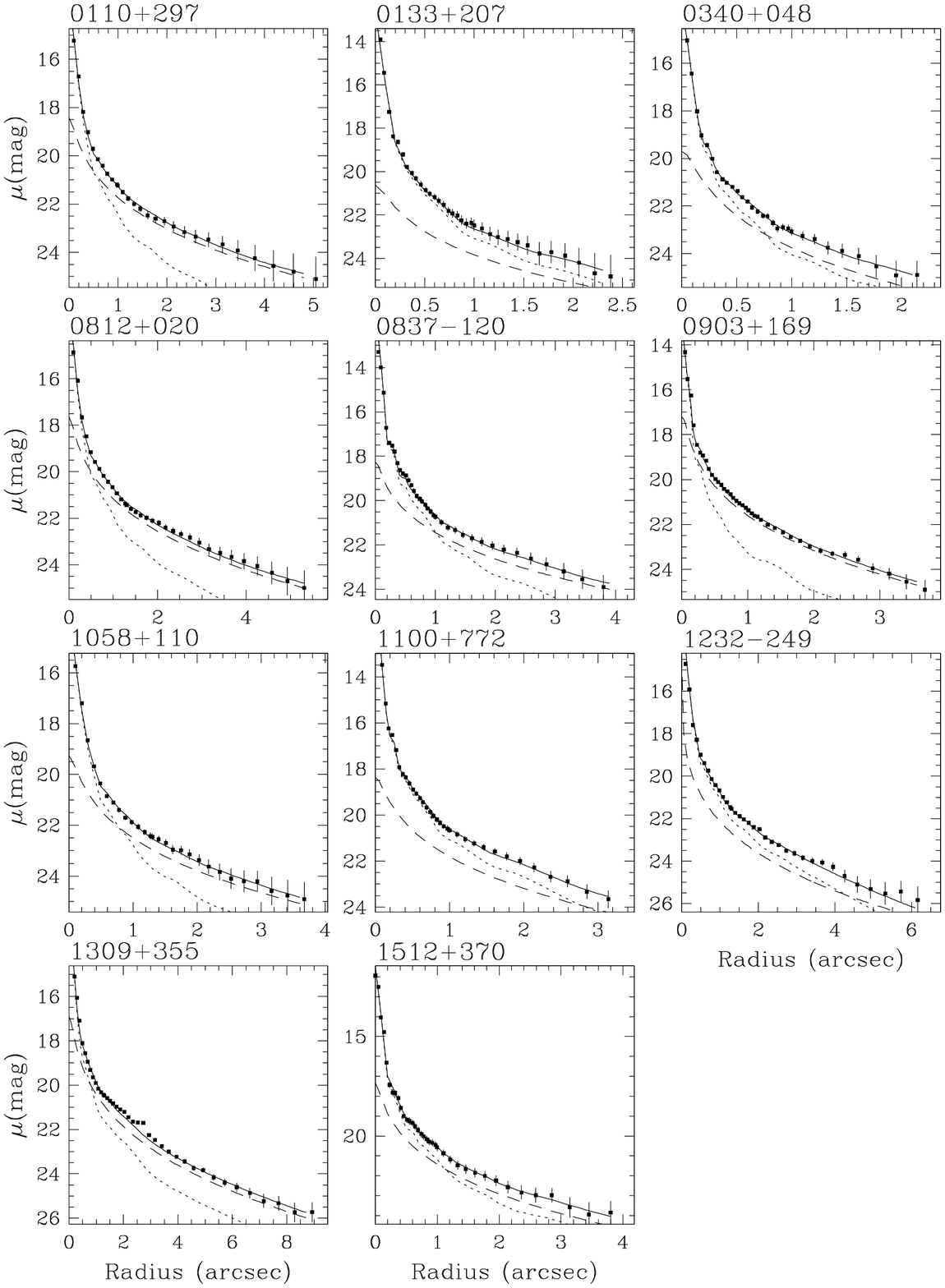}
\caption{The observed radial brightness profiles of the quasars
(filled squares), compared with the best fitted model (solid line) consisting of 
a point source modeled by a scaled 
PSF (dotted line) and the elliptical (de Vaucouleurs) galaxy
convolved with its PSF (dashed line).}
\end{figure}

% FIG 5 - Mhost and Re distribution and Mnucl distribution
%
%\begin{figure}[h]
%\plotthree{hstqso_fig5a.ps}{hstqso_fig5b.ps}{hstqso_fig5c.ps}
%\caption{ Distribution of host galaxy properties 
%(absolute magnitudes and effective radii) of the full sample of 34 RLQ}
%\end{figure}
\begin{figure}[h]
\epsscale{0.45}
\plotone{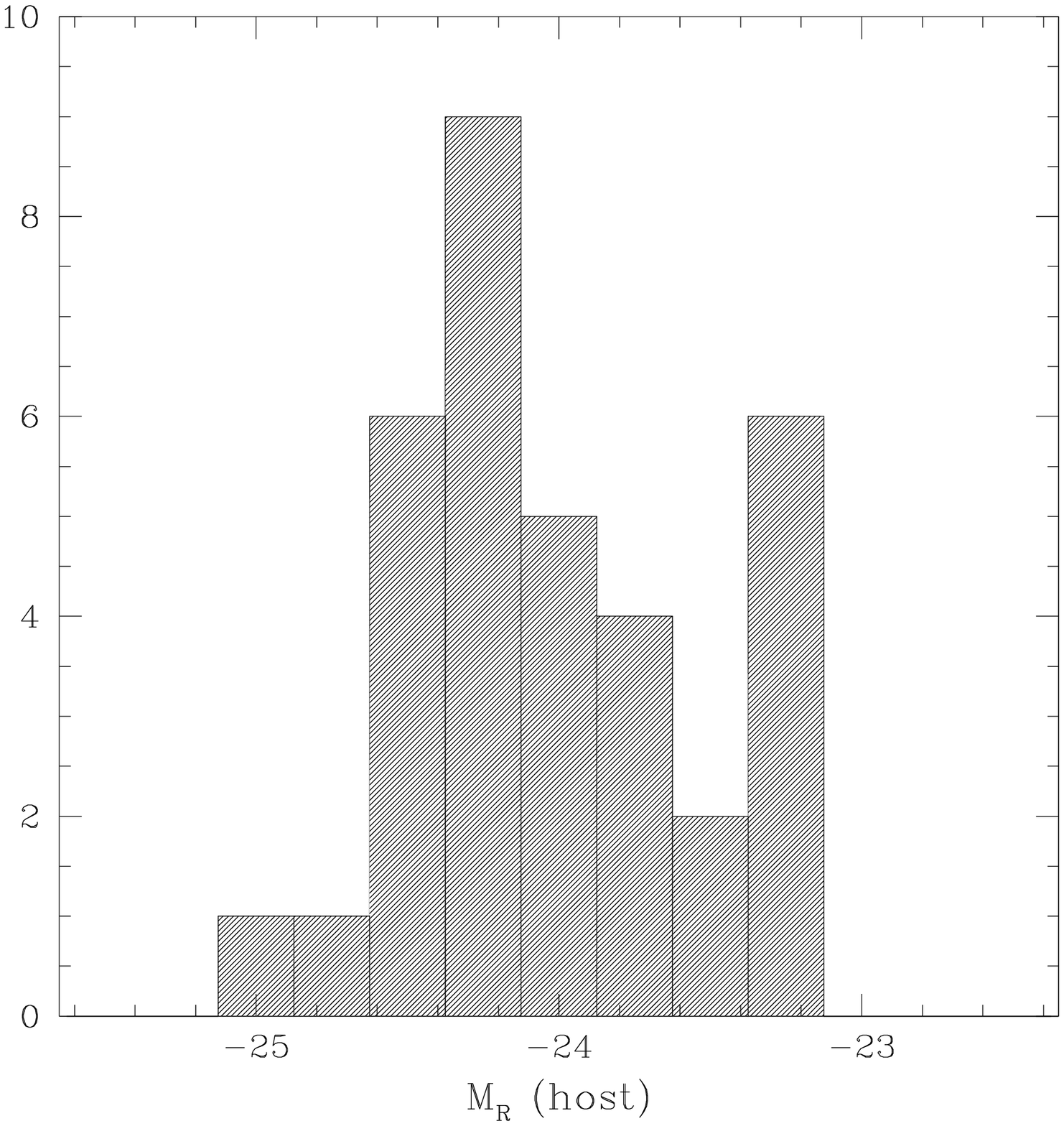}\\
\epsscale{0.45}
\plotone{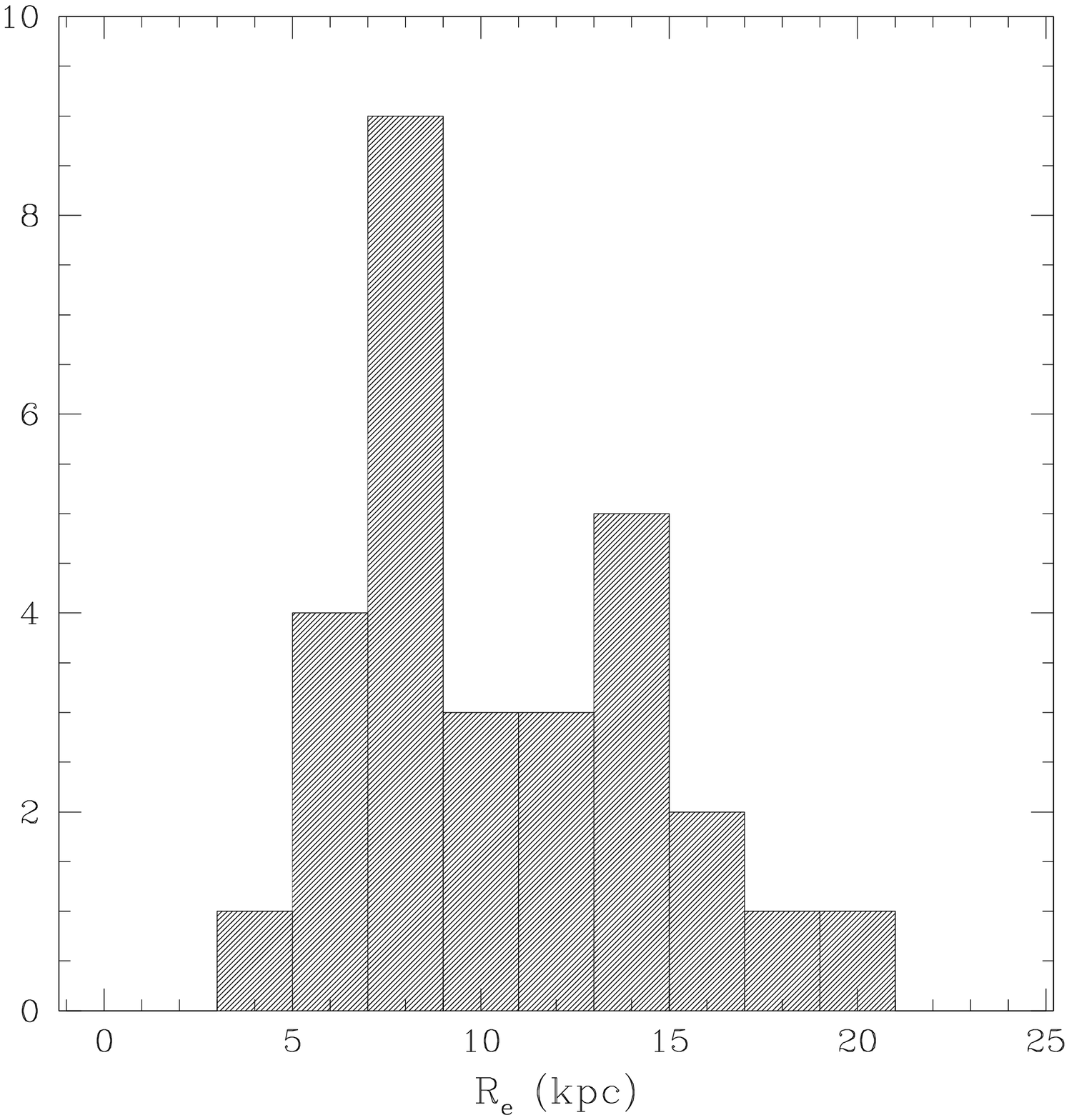}\\
\epsscale{0.45}
\plotone{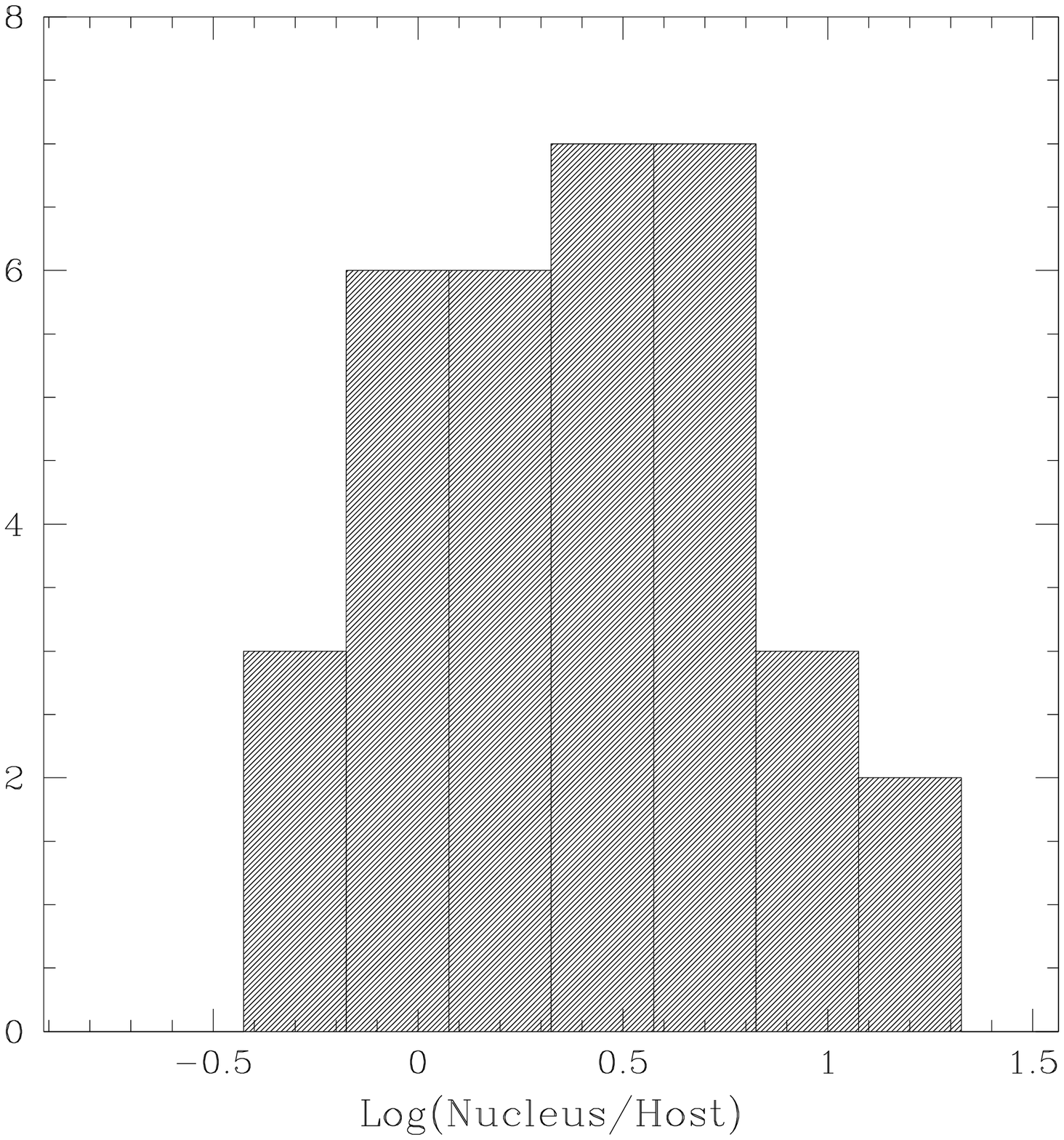}
\caption{Distribution of host galaxy properties 
(absolute magnitudes ({\it top}) and effective radii (center)) 
and of the ratio of quasar to galaxy
luminosity of the full sample of 34 RLQ}
\end{figure}

% FIG 6  Mhost Mnuc diagram
%
\begin{figure}[h]
%\centerline{\mbox{\epsfysize=12.0truecm\epsffile{mhmn.ps}}}
\epsscale{1}
\plotone{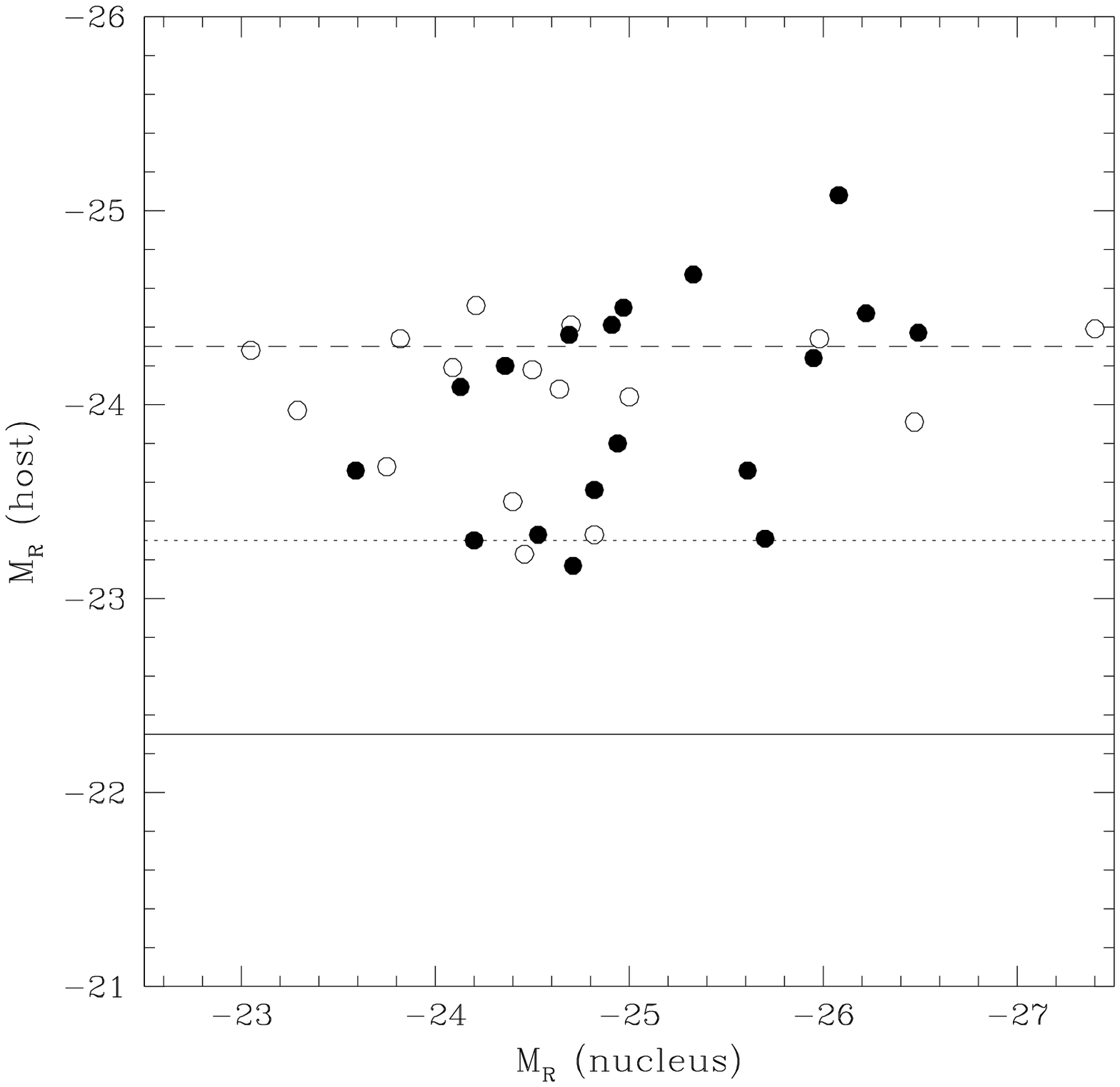}
\caption{The QSO host galaxy absolute magnitudes compared with the 
 nuclear luminosity (R band).
For the majority of the objects the host luminosity is encompassed 
between 2L$^*$ and 5L$^*$ while 
nuclear luminosities differ by a factor 40. 
No significant difference is found between objects at 
z $<$ 0.3 (open symbols) and those at z $>$ 0.3. 
Horizontal lines represent the luminosity of an M$^*$ (solid line) 
,M$^*$-1 (dotted line), and M$^*$-2 (dashed line) at z=0.}
\end{figure}

% FIG 7  Mbh - Log(Lum) and Ledd
%
\begin{figure}
\plotone{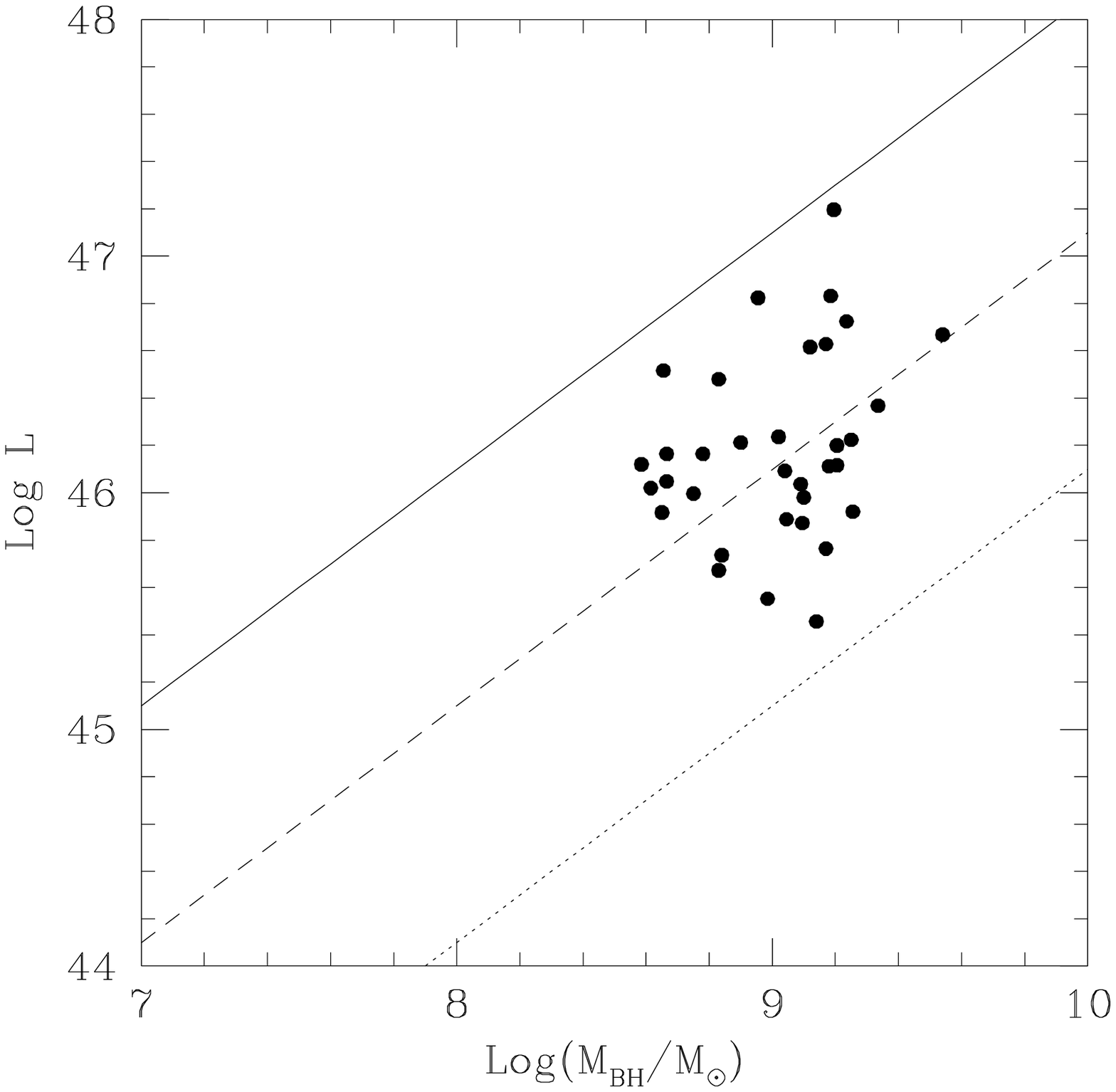}
\caption{The black hole mass of the RLQ derived from the host luminosity (see text) 
compared with the 
bolometric luminosity of the objects derived from the observed (R-band) nuclear 
luminosities after applying bolometric correction (Laor and Draine 1993).
The diagonal lines represent the expected position for the objects is they are radiating at 
their respective Eddington luminosity (solid line) and at 10\% and 1\% of L$_{Edd}$ 
(dashed line and dotted line, respectively). 
A large spread of Eddington ratio is apparent which 
is not explained by the uncertainity in the estimates of $\CMcal M_{BH}$. }
\end{figure}

% FIG 8  Mbh radio-power
%
\begin{figure}
\plotone{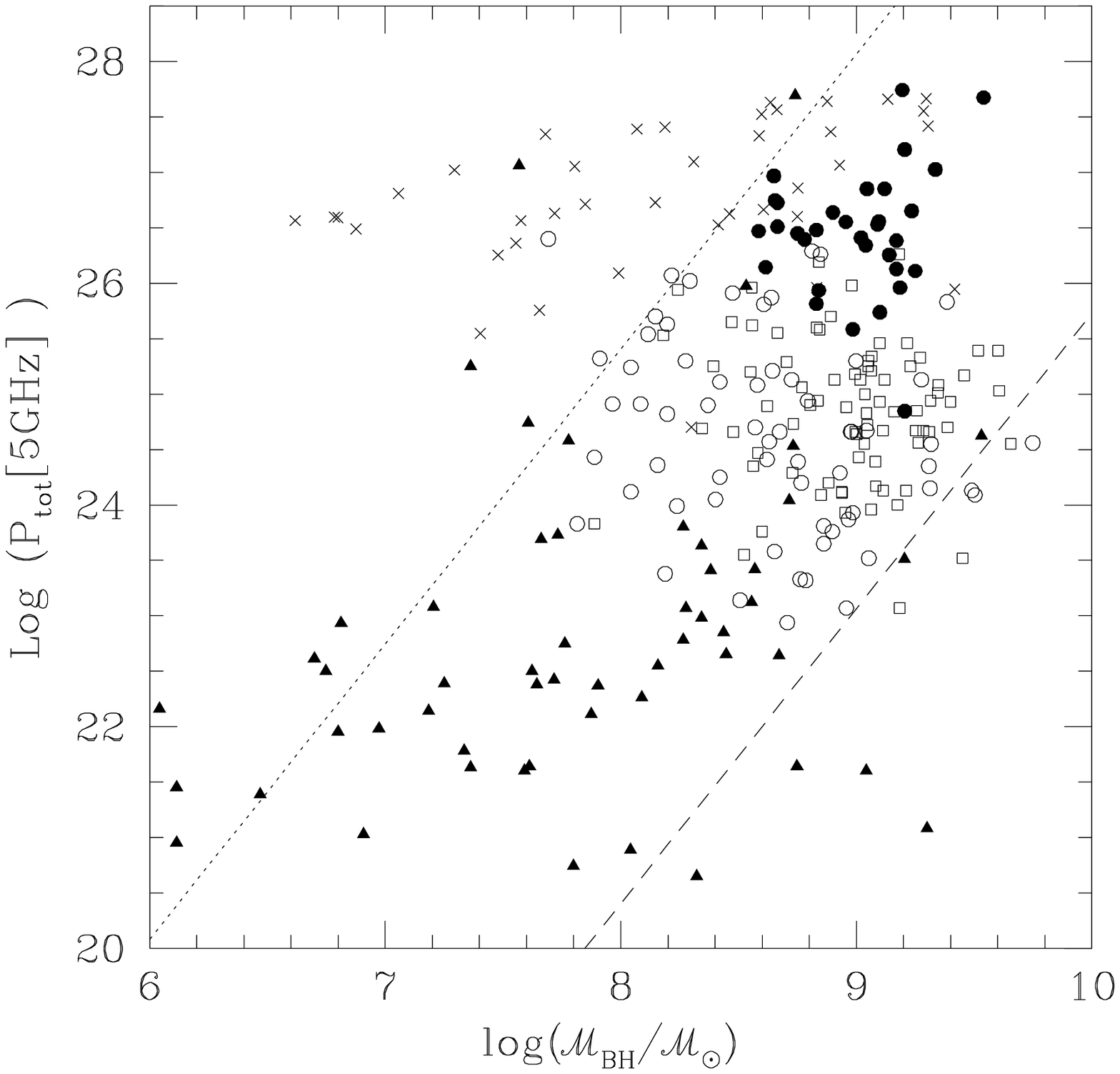}
\caption{$\CMcal M_{BH}$-LogP(total) relation for the 34 RLQ (filled circles)
compared with the sample of radiogalaxies (open circle and
squares) from \citealp{bettoni}, with
the objects studied by Ho (2002) (triangles) and
with FSRQ (crosses) investigated by Oshlack et
al. (2002). The dashed line is the Franceschini et al. 1996 relation
($P_{tot}\propto \CMcal M_{BH}^{2.5}$) while the dotted line corresponds to an 
offset of 5 order of magnitude. 
Note that the points that deviate from the region encompassed by these two lines 
refer mainly to FSRQ and could be reconciled with the overall trend 
if corrections for beaming and orientation are taken into account (see text)}
\label{Fig8}
\end{figure}

\end{document}